# Structural, elastic and electronic properties of Fe$_3$C from first-principles

C. Jiang[1,*], S. G. Srinivasan[1], A. Caro[2], and S. A. Maloy[1]

[1]Materials Science & Technology Division, Los Alamos National Laboratory, Los Alamos, NM 87545

[2]Materials Science & Technology Division, Lawrence Livermore National Laboratory, Livermore, CA 94550

Using first-principles calculations within the generalized gradient approximation, we predicted the lattice parameters, elastic constants, vibrational properties, and electronic structure of cementite (Fe$_3$C). Its nine single-crystal elastic constants were obtained by computing total energies or stresses as a function of applied strain. Furthermore, six of them were determined from the initial slopes of the calculated longitudinal and transverse acoustic phonon branches along the [100], [010] and [001] directions. The three methods agree well with each other, the calculated polycrystalline elastic moduli are also in good overall agreement with experiments. Our calculations indicate that Fe$_3$C is mechanically stable. The experimentally observed high elastic anisotropy of Fe$_3$C is also confirmed by our study. Based on electronic density of states and charge density distribution, the chemical bonding in Fe$_3$C was analyzed and was found to exhibit a complex mixture of metallic, covalent, and ionic characters.

*Corresponding author: chao@lanl.gov

## I. INTRODUCTION

Iron and steels continue to be the most important structural material several millennia after their discovery. Precipitation of excess carbon as cementite ($Fe_3C$) instead of graphite in the form of pearlite, upon slow cooling to below 1000K, is ubiquitous in steels with carbon content greater than 0.02%. Cementite is important because its morphology directly controls the mechanical properties of steels. Contrasting to its technological significance, knowledge of the elastic properties of cementite, critical to understanding the mechanical properties of steels, is quite limited. Until recently, polycrystalline bulk cementite was successfully synthesized through a combination of mechanical alloying and spark plasma sintering.[1,2] To date, direct experimental measurements of the single-crystal elastic constants of cementite are still not feasible due to difficulties in growing large enough single-crystal samples. From the theoretical point of view, the electronic structure of cementite has been calculated by various authors.[3-5] Its bulk modulus has also been recently calculated by Faraoun *et al.*,[5] Vocadlo *et al.*,[6] and Huang *et al.*[7]

In this work, we present first-principles calculated results for various physical properties of cementite, including its lattice parameters, the complete set of single-crystal elastic constants, equation of state, and phonon dispersion relations. It is also our purpose to provide some insight into the nature of bonding in cementite by investigating its electronic structure. Aside from the practical importance of our results, they can be used by coarser scale approaches[8] to develop a predictive computational model for steels.

## II. COMPUTATIONAL DETAILS

First-principles calculations were performed using projector augmented wave (PAW)[9] pseudopotentials and a plane-wave basis set, as implemented in Vienna *ab initio* simulation package (VASP).[10] The exchange-correlation functional was described within the generalized gradient approximation (GGA) as parameterized by Perdew, Burke, and Ernzerhof (PBE).[11] It is worth noting that GGA correctly predicts the ferromagnetic bcc structure of Fe as its ground state, while the local density approximation incorrectly predicts its ground state to be a nonmagnetic close-packed structure.[12, 13] Spin-polarized calculations were performed to model the ferromagnetically ordered cementite. The *k*-point meshes for Brillouin zone sampling were constructed using the Monkhorst–Pack scheme.[14] After careful convergence tests, a plane wave cutoff energy of 500 eV and an 11×9×13 *k*-point mesh for the 16-atom unit cell of cementite were found to be sufficient to converge the elastic constants to ~1GPa and the total energy to better than 1meV/atom.

## III. RESULTS AND DISCUSSION

### A. Structural properties

Cementite has a complex orthorhombic crystal structure (space group Pnma, No. 62) with 12 Fe atoms and 4 C atoms per unit cell.[15, 16] The 12 Fe atoms are distributed between two distinct lattice sites: the special $Fe_I$ sites (Wyckoff position 4*c*) and the general $Fe_{II}$ sites (Wyckoff position 8*d*). As shown in Fig. 1, the fundamental building block of cementite is the trigonal prism formed by a C atom and its six nearest-neighbor Fe atoms (2 $Fe_I$ + 4 $Fe_{II}$). The trigonal prisms are interconnected through edge- and corner-sharing to form a layer. A stacking of those layers along the *b* axis thus forms the cementite structure. By computing the quantum mechanical forces and stress tensor, we fully relaxed the unit cell volume and shape as well as all internal atomic coordinates of cementite using a conjugate-gradient scheme, and our results are summarized in Table I. The lattice constants calculated within the GGA agree with the experimental data from Wood *et al.*[16] to within 1%. The computed cell-internal parameters are also in excellent agreement with experiments. These results give us confidence in the accuracies of first-principles calculations in predicting the properties of cementite.

**B. Single-crystal elastic constants**

An orthorhombic crystal has nine independent single-crystal elastic constants. They can be determined by applying a small strain to the equilibrium lattice and computing the resultant change in its total energy.[17, 18] The distorted lattice vectors can be obtained via a matrix multiplication $R' = RD$, where $R'$ and $R$ are the 3×3 matrices containing the components of the distorted and undistorted lattice vectors, respectively. $D$ is a symmetric distortion matrix:

$$D = \begin{pmatrix} 1+\varepsilon_1 & \varepsilon_6/2 & \varepsilon_5/2 \\ \varepsilon_6/2 & 1+\varepsilon_2 & \varepsilon_4/2 \\ \varepsilon_5/2 & \varepsilon_4/2 & 1+\varepsilon_3 \end{pmatrix} \qquad (1)$$

Where $\varepsilon_i$ is the *i*-th component of the strain tensor in Voigt notation (*xx*=1, *yy*=2, *zz*=3, *yz*=4, *xz*=5, and *xy*=6). Here indices 1 to 3 are associated with normal strains (or stresses), and 4 to 6 are associated with shear components.

In the framework of linear elasticity theory, the distortion energy can be written as a function of strain as:

$$\Delta E = E - E_0 = \frac{V_0}{2}\sum_{i=1}^{6}\sum_{j=1}^{6}c_{ij}\varepsilon_i\varepsilon_j + O(\varepsilon^3) \qquad (2)$$

Where $E$ and $E_0$ are the total energies of the distorted and equilibrium structures, respectively. $V_0$ is the equilibrium volume. The nine distortion matrices necessary for computing the nine elastic constants of an orthorhombic structure are given in Table II. At each given strain, we fully relaxed all atoms to their energetically more favorable lattice positions according to strain-induced forces while holding the distorted lattice vectors fixed. We then fitted the distortion energies calculated at $\delta=\pm0.005$ and $\pm0.01$ to the function $\Delta E(\delta) = k_2\delta^2 + k_3\delta^3$, and obtained the single-crystal elastic constants $c_{ij}$ from the quadratic coefficient $k_2$ (see Table II for relationships).

For the purpose of comparison, we also obtained the $c_{ij}$ of cementite using a stress-strain approach based on the generalized Hooke's law.[19, 20] For a small strain $\varepsilon$, the corresponding stress components $\sigma_i$ ($i$=1 to 6) can be written as a linear function of strain as:

$$\sigma_i = \sum_{j=1}^{6} c_{ij} \varepsilon_j \tag{3}$$

The elastic constants of cementite can thus be extracted from a linear least-square fit of the first-principles Hellmann-Feynman stresses as a function of strain. Here only six distortion matrices (1-3 and 7-9 in Table II) are needed to compute all nine elastic constants. Thus the stress-strain approach is computationally more efficient than the energy-strain approach, especially for structures with low symmetry.

Table III reports our calculated single-crystal elastic constants of cementite. The energy-strain and stress-strain approaches agree rather well with each other. One condition for mechanical stability of a structure is that its strain energy must be positive against any homogeneous elastic deformation. For an orthorhombic crystal, this imposes the following constraints: $c_{11} + c_{22} > 2c_{12}$, $c_{22} + c_{33} > 2c_{23}$, $c_{11} + c_{33} > 2c_{13}$, $c_{ii} > 0$ ($i$=1 to 6), and $c_{11} + c_{22} + c_{33} + 2c_{12} + 2c_{23} + 2c_{13} > 0$.[21] The fact that the elastic constants of cementite obey all the above criteria indicates that it is an elastically stable structure.

In Table III, the results from unrelaxed calculations, in which the internal atomic positions were "frozen" at their zero-strain equilibrium values after the unit cell shape is deformed, are also shown. It can be seen that internal atomic relaxations reduce $c_{22}$ and $c_{33}$ by about 68 and 56 GPa, respectively, even though the corresponding distortions preserve the orthorhombic symmetry of cementite. This can be understood from the fact that even the unstrained cementite structure contains many internal atomic coordinates not fixed by space group symmetry. Under a given strain, those internal atomic degrees of freedom (DOF) will be re-adjusted to minimize the total energy of the system associated with that strain. Since such a process will always lower the energy, it will always reduce the elastic constants that appear as the sole coefficient in the energy vs. strain relations. For $c_{12}$, $c_{23}$, and $c_{13}$, they can never appear isolated but have to occur in combination with other elastic constants in the energy vs. strain relations.[22] As the result, those elastic constants need not decrease upon internal relaxation.

In addition to the DOF already available in the unstrained cementite, new DOF can be introduced as the result of strain-induced symmetry reduction, *e.g.* from orthorhombic to monoclinic upon shear deformation (see Table II). For the strain corresponding to shear modulus $c_{44}$, the symmetry-reduction allows the *y*-coordinates of $Fe_I$ and C atoms, which are initially fixed by symmetry in the unstrained unit cell, to relax. In fact, we found that C atoms adjust to the shear by moving mainly along the *b* axis. The corresponding dimensionless Kleinman internal displacement parameter was calculated to be a large value of 0.22. As the consequence of such a large relaxation, the elastic constant $c_{44}$, which measures the resistance of the lattice against (010)[001] shear, is reduced from ~82 GPa to an anomalously low value of ~15 GPa. We note that, the strong relaxation-induced softening of elastic constants has also been observed by Panda and Chandran[17, 18] in $TiB_2$ and TiB and by Wang *et al.*[23] in $Al_3BC_3$. It has also been demonstrated by Lu and Klein[24] that, for compounds $V_3Si$, $V_3Ge$ and $Nb_3Sn$, the effects of internal relaxations are even significant enough to make their shear elastic constants negative.

**C. Vibrational properties**

In addition to the elastic stability criterion, dynamical stability further requires that the phonon spectra of a structure exhibit no imaginary phonon frequencies, or the so-called soft modes.[25, 26] A soft mode indicates instability with respect to a particular type of atomic movement, thus signals structural phase transitions. To calculate the phonon spectra of $Fe_3C$, we used the direct approach[26-28] as implemented in ATAT.[29] In this approach, the 3×3 interatomic force constant matrices were fitted to Hellmann-Feynman forces on all atoms in perturbed supercells. From the fitted force constants, one can construct a dynamical matrix for arbitrary wave vector in the Brillouin zone and obtain the phonon frequencies by solving the eigenvalue problem. In our phonon calculations, we employed large 128-atom supercells to allow inclusion of long-range (up to 5.5Å) force constants in the fitting. The magnitude of atomic displacements was chosen to be 0.05 Å. Our calculated phonon dispersion relations for cementite along various high-symmetry directions in the Brillouin zone are shown in Fig. 2(a). There are 48 phonon branches since the unit cell of cementite contains 16 atoms. It is evident that cementite is also dynamically stable since all phonon frequencies are positive throughout the Brillouin zone.

The low-frequency part of the calculated phonon spectra allows us to determine the sound wave velocities propagating through the crystal, which are in turn related to the single-crystal elastic constants via the Christoffel equation.[30] These relationships for an orthorhombic crystal are given in Table IV. The elastic constants derived from the slopes of the long wavelength acoustic phonon branches along [100], [010] and [001] directions are also shown in Table III and they agree well with our directly calculated elastic constants. It is worth noting that the two low-lying TA branches along the [010] and [001] directions are a manifestation of the anomalously soft elastic shear modulus $c_{44}$. Also, the two TA branches along the [100] direction are degenerate in the long-wavelength limit because of the similar values of $c_{55}$ and $c_{66}$.

Finally, to test the accuracy of GGA, we have also calculated the phonon spectra of diamond and ferromagnetic bcc Fe, where experimental neutron diffraction measurements[31-33] are available. As shown in Fig. 2(b) and 2(c), the agreement between experiments and calculations is quite satisfactory for both diamond and Fe. The only large discrepancy exists at the N point for bcc Fe. To verify this, the frozen-phonon method[34] was employed to calculate the phonon frequencies at three high-symmetry $k$-points in the Brillouin zone: $H=\frac{2\pi}{a}(0,0,1)$, $N=\frac{2\pi}{a}(\frac{1}{2},\frac{1}{2},0)$, and $P=\frac{2\pi}{a}(\frac{1}{2},\frac{1}{2},\frac{1}{2})$. In this approach, one calculates the total energy of a system as a function of the amplitude of a "frozen" phonon wave, and the phonon frequencies can thus be directly calculated from the curvature of the energy vs. amplitude curve. The phonon frequencies thus obtained can be considered as "exact" since they do not depend on the interaction range of force constants. Our calculated phonon frequencies at N, P, and H points are also shown in Fig. 2(c), in good agreement with our direct force constant calculations. We thus conclude that the discrepancy at the N point is not due to truncation of the force constant ranges in the direct method, but is rather due to the intrinsic error of GGA.

**D. Polycrystalline elastic moduli**

As discussed, experimental single-crystal elastic constants are not available for cementite. However, polycrystalline elastic moduli of cementite have been experimentally measured and can be directly compared with our calculations. By treating the polycrystalline materials as aggregates of single crystals with random orientations, the isotropic polycrystalline elastic moduli can be computed as averages of anisotropic single-crystal elastic constants. The theoretical lower and upper bounds to the true polycrystalline bulk modulus $B$ and shear modulus $G$ are given by Reuss (assuming uniform stress throughout a polycrystal) and Voigt (assuming uniform strain) as follows:[18]

$$B_R = \frac{1}{s_{11} + s_{22} + s_{33} + 2(s_{12} + s_{23} + s_{13})} \tag{4a}$$

$$B_V = \frac{c_{11} + c_{22} + c_{33} + 2(c_{12} + c_{23} + c_{13})}{9} \tag{4b}$$

$$G_R = \frac{15}{4(s_{11} + s_{22} + s_{33} - s_{12} - s_{23} - s_{13}) + 3(s_{44} + s_{55} + s_{66})} \tag{4c}$$

$$G_V = \frac{c_{11} + c_{22} + c_{33} - c_{12} - c_{23} - c_{13}}{15} + \frac{c_{44} + c_{55} + c_{66}}{5} \tag{4d}$$

Where $s_{ij}$ are elastic compliances and their values can be obtained through an inversion of the elastic constant matrix, $S = C^{-1}$. We estimated the true polycrystalline values by the Voigt-Reuss-Hill approximation: $B = (B_R + B_V)/2$ and $G = (G_R + G_V)/2$.[35] The Young's modulus and Poisson's ratio can then be calculated as $Y = 9BG/(3B+G)$ and $\upsilon = (3B/2 - G)/(3B+G)$.

Our calculated polycrystalline elastic moduli of cementite are reported in Table V, in comparison with available experiments[2, 36-41] and previous theoretical calculations.[5-7] The overall agreement is remarkable. The only exception is the bulk modulus: although our calculated values agree well with previous theoretical calculations, they are ~28% larger than those obtained from high-pressure diamond anvil cell experiments.[36-38] This is however atypical of GGA, which usually underestimates the elastic constants. It is worth point out that, our bulk modulus was calculated from single-crystal elastic constants obtained strictly within the linear elasticity regime of the material (*i.e.*, from small strains around the equilibrium state), while the experimental bulk moduli were obtained from an equation of state (EOS) fit of measurements at high pressures, where nonlinear elasticity effects are significant. To reproduce the experimental conditions, we calculated the total energy of cementite at 10 different volumes corresponding to a wide range of pressures (0-30 GPa) and fitted the obtained *E-V* data to an integrated third-order Birch-Murnaghan[42] EOS. At each volume, the unit cell shape (*b/a* and *c/a* ratios) and all internal atomic positions were fully relaxed. The pressure at any volume can then be calculated as $P = -\partial E/\partial V$. Our first-principles EOS is directly compared with the experimental measurements[36-38] in Fig. 3. Compared to experiments, our calculations slightly underestimate the decrease of volume at high pressures. Our non-linear fitting gives the zero-pressure bulk modulus $B_0$ of cementite to be 204 GPa, a value in much better agreement with experiments. The fitted pressure derivative of bulk modulus (5.0) also agrees well with experiments. It is worth noting that we have also performed an EOS fit of a separate set of *E-V* data corresponding to a pressure range from about -3 to 6 GPa, which are close to the equilibrium state. Our fitting now gives $B_0$ to be 223 GPa, in good agreement with the bulk modulus from Hill's average. Our study thus suggests that the bulk modulus obtained from an EOS fit can be dependent on the pressure range used in the fit. As the consequence, an EOS fitted to high-pressure experimental measurements may not necessarily be accurate at low pressures, and vice-versa.

**E. Elastic anisotropy**

All crystals exhibit varying degrees of elastic anisotropic behavior. The elastic anisotropy of an orthorhombic crystal can be measured by three shear anisotropy factors (Zener ratios): $A_1 = 4c_{44}/(c_{22} + c_{33} - 2c_{23})$, $A_2 = 4c_{55}/(c_{11} + c_{33} - 2c_{13})$, and $A_3 = 4c_{66}/(c_{11} + c_{22} - 2c_{12})$ for the (100), (010) and (001) shear planes, respectively. The calculated values for cementite are shown in Table VI. For an isotropic crystal, all three factors must be one. Our results thus indicate a very large shear anisotropy on the (100) plane of cementite due to the anomalously low $c_{44}$. In addition to shear anisotropy, one also needs to consider the anisotropy in linear bulk modulus for an orthorhombic crystal.

For an orthorhombic crystal, the linear bulk modulus in an arbitrary direction can be calculated from the elastic compliances as:[43]

$$\frac{1}{B} = (s_{11} + s_{12} + s_{13})l_1^2 + (s_{12} + s_{22} + s_{23})l_2^2 + (s_{13} + s_{23} + s_{33})l_3^2 \qquad (5)$$

where $l_1$, $l_2$, and $l_3$ are direction cosines. Our calculated linear bulk moduli of cementite along $a$, $b$, and $c$ axes are given in Table VI. For an isotropic system, the three linear bulk moduli must be equal. Table VI shows that the linear bulk modulus of cementite is the smallest along the $c$ axis, indicating that axis to be the most compressible. The elastic anisotropy of a crystal can also be measured by the percentage anisotropy in compressibility and shear: $A_B = (B_V - B_R)/(B_R + B_V)$ and $A_G = (G_V - G_R)/(G_R + G_V)$.[18] A value of 0 represents elastic isotropy and a value of 1 represents the maximum anisotropy. As can be seen from Table VI, cementite exhibits a high degree of anisotropy in shear but only a small anisotropy in compressibility.

Experimentally, it is known that the Young's modulus of cementite is highly anisotropic, and its Young's modulus in the [100] direction ($a$ axis) is larger than in the directions normal to [100].[44] For an orthorhombic crystal, the Young's modulus in an arbitrary direction can be calculated as:[43]

$$\frac{1}{Y} = l_1^4 s_{11} + l_2^4 s_{22} + l_3^4 s_{33} + l_1^2 l_2^2 (2s_{12} + s_{66}) + l_2^2 l_3^2 (2s_{23} + s_{44}) + l_1^2 l_3^2 (2s_{13} + s_{55}) \qquad (6)$$

Fig. 4 illustrates the directional dependence of the Young's modulus of cementite calculated using the singe-crystal elastic constants from the present study. For an isotropic system, one would see a spherical shape. In agreement with experiments, Fig. 4 shows a considerable deviation in shape from a sphere, thus indicating a high degree of anisotropy in Young's modulus. Remarkably, Fig. 4 also directly confirms the experimental observation by Kagawa et al.[44] that the Young's modulus of cementite along the *a* axis is larger than in the directions normal to it.

**F. Nature of bonding in cementite**

To gain some insight into the nature of bonding in cementite, we have also investigated its electronic structure. Fig. 5 presents the partial electronic density of states (DOS) of cementite. The general features agree well with the previous calculations by Haglund et al.,[3] Shein et al.,[4] and Faraoun et al.[5] Cementite is metallic because of the finite DOS at the Fermi level. The lowest-lying states are mainly from C 2s states, while the DOS near the Fermi level are all dominated by Fe 3d states. The large asymmetry of Fe 3d states is a strong indication of local magnetic moment. We calculated the magnetic moment to be 1.96 1.89, and -0.12 $\mu_B$ at the $Fe_I$, $Fe_{II}$, and C sites, respectively. Such values are in good agreement with previous calculations.[3-5] Importantly, Fig. 5 shows a hybridization between C 2p and Fe 3d states in the energy range from -8 eV to -4 eV below the Fermi level, an evidence of covalent bonding between Fe and C atoms. The corresponding anti-bonding states lie above the Fermi level.

Fig. 6(a) further shows the charge density distribution in the (010) plane of cementite. It can be seen that the bonding between Fe and C exhibits some directionality, indicative of covalent bonding. To investigate the charge transfer, we also plotted in Fig. 6(b) the difference in charge density between $Fe_3C$ compound and superposition of free atoms in the same plane. There is significant charge depletion at the Fe atoms. Some of those electrons become delocalized to form "a sea of electrons", typical of metallic bonding. Some of them are transferred to the C atoms, revealing an ionic contribution to the Fe-C bonding. The predicted charge transfer direction is also in agreement with our Bader charge density analysis,[45] and the fact that C has a larger electronegativity (2.55) than Fe (1.8). We thus conclude that the bonding in cementite has a complex mixture of metallic, covalent, and ionic characters.

Finally, elastic constants may provide some insight into the bonding. As discussed, the cementite structure can be geometrically described as a stacking of layers formed by interconnected trigonal prisms along the $b$ axis. Within each layer, there exist a mixture of Fe-C and Fe-Fe bonding. Nevertheless, coupling between adjacent layers is predominantly due to Fe-Fe bonding (see Fig. 1). Assuming that the cementite structure is dominated by Fe-C bonding, one would expect the $b$ axis to be the weakest. This is however not the case since $c_{22}$ is in fact larger than $c_{33}$ (see Table III). Clearly, Fe-Fe bonding also plays an important role in cementite. It is worth noting that, such a conclusion is in agreement with recent calculations by Medvedeva et al.[46] showing that Fe-C bonding only accounts for ~38% of the cohesive energy of $Fe_3C$.

## IV. CONCLUSIONS

Motivated by lack of experiments, we have performed a theoretical study of the structural, elastic and electronic properties of cementite using first-principles calculations. The calculated equilibrium lattice parameters and internal atomic coordinates are in excellent agreement with experiments. The full set of single-crystal elastic constants of cementite calculated using three different approaches are in good agreement with each other. Our results indicate that $Fe_3C$ is mechanically stable and exhibits a high degree of anisotropy. Our predicted anisotropy in Young's modulus agrees with experimental measurements. By averaging single-crystal elastic constants, the polycrystalline elastic moduli of $Fe_3C$ were also calculated and the results are in overall good agreement with experiments, suggesting that our single-crystal elastic constants are also accurate. We hope that our calculations will stimulate further experimental study on this important compound.

## ACKNOWLEDGEMENTS

At Los Alamos National Laboratory (LANL), we acknowledge the support of Director's postdoctoral fellowship and the Global Nuclear Energy Partnership (GNEP) program (USDOE-NE). Work of AC was performed under the auspices of the U.S. Department of Energy by the University of California, Lawrence Livermore National Laboratory under Contract No. W-7405-Eng-48. We also thank John Wills (LANL, Theoretical Division) and M.I. Baskes (LANL, Materials Science Division) for their many helpful suggestions.

**TABLE I**. Experimental and theoretical structural properties of cementite. $(x, y, z)_i$ denotes the internal atomic coordinates of atom $i$.

| Property | Experiment[16] | Present Study |
|---|---|---|
| Lattice constants (Å) | $a$=5.08, $b$=6.73, $c$=4.51 | $a$=5.04, $b$=6.72, $c$=4.48 |
| $(x, y, z)_{FeI}$ | (0.034,0.25,0.841) | (0.036,0.25,0.837) |
| $(x, y, z)_{FeII}$ | (0.184,0.057,0.333) | (0.176,0.068,0.332) |
| $(x, y, z)_C$ | (0.894,0.25,0.450) | (0.876,0.25,0.438) |

**TABLE II.** Deformation matrices used to calculate the nine independent single-crystal elastic constants of an orthorhombic crystal.

| Number | Strain (all unlisted $\varepsilon_i$=0) | Crystal structure after deformation | Relationship between $k_2$ and elastic constants |
|---|---|---|---|
| 1 | $\varepsilon_1=\delta$ | Orthorhombic | $k_2 = \frac{1}{2}V_0 C_{11}$ |
| 2 | $\varepsilon_2=\delta$ | Orthorhombic | $k_2 = \frac{1}{2}V_0 C_{22}$ |
| 3 | $\varepsilon_3=\delta$ | Orthorhombic | $k_2 = \frac{1}{2}V_0 C_{33}$ |
| 4 | $\varepsilon_1=\delta$, $\varepsilon_2=-\delta$ | Orthorhombic | $k_2 = \frac{1}{2}V_0 (C_{11}+C_{22}-2C_{12})$ |
| 5 | $\varepsilon_2=\delta$, $\varepsilon_3=-\delta$ | Orthorhombic | $k_2 = \frac{1}{2}V_0 (C_{22}+C_{33}-2C_{23})$ |
| 6 | $\varepsilon_1=\delta$, $\varepsilon_3=-\delta$ | Orthorhombic | $k_2 = \frac{1}{2}V_0 (C_{11}+C_{33}-2C_{13})$ |
| 7 | $\varepsilon_4=2\delta$ | Monoclinic | $k_2 = 2V_0 C_{44}$ |
| 8 | $\varepsilon_5=2\delta$ | Monoclinic | $k_2 = 2V_0 C_{55}$ |
| 9 | $\varepsilon_6=2\delta$ | Monoclinic | $k_2 = 2V_0 C_{66}$ |

**TABLE III.** Single-crystal elastic constants (in GPa) of cementite calculated in the present study. The values shown in parentheses are from unrelaxed calculations, *i.e.*, without allowing for internal atomic relaxations.

| Method | $c_{11}$ | $c_{22}$ | $c_{33}$ | $c_{12}$ | $c_{23}$ | $c_{13}$ | $c_{44}$ | $c_{55}$ | $c_{66}$ |
|---|---|---|---|---|---|---|---|---|---|
| Energy-strain | 388 (413) | 345 (412) | 322 (378) | 156 (154) | 162 (170) | 164 (167) | 15 (82) | 134 (136) | 134 (140) |
| Stress-strain | 395 (417) | 347 (416) | 325 (381) | 158 (157) | 163 (174) | 169 (171) | 18 (82) | 134 (136) | 135 (140) |
| Phonon | 384 | 325 | 283 | - | - | - | 26 | 134 | 125 |

**TABLE IV.** The sound velocities in an orthorhombic crystal. $\rho$ denotes the density of the material.

| Direction | Mode | Polarization | Sound Velocity |
|---|---|---|---|
| [100] | Longitudinal | [100] | $\sqrt{c_{11}/\rho}$ |
| [100] | Transverse | [010] | $\sqrt{c_{66}/\rho}$ |
| [100] | Transverse | [001] | $\sqrt{c_{55}/\rho}$ |
| [010] | Longitudinal | [010] | $\sqrt{c_{22}/\rho}$ |
| [010] | Transverse | [100] | $\sqrt{c_{66}/\rho}$ |
| [010] | Transverse | [001] | $\sqrt{c_{44}/\rho}$ |
| [001] | Longitudinal | [001] | $\sqrt{c_{33}/\rho}$ |
| [001] | Transverse | [100] | $\sqrt{c_{55}/\rho}$ |
| [001] | Transverse | [010] | $\sqrt{c_{44}/\rho}$ |

**TABLE V.** Polycrystalline elastic moduli (in GPa) of cementite obtained from first-principles calculations and experiments. Experimental values of the first pressure derivative of bulk modulus are shown within parentheses.

| Property | Present Study | | Previous Calculations | Experiments |
|---|---|---|---|---|
| | Energy-strain | Stress-strain | | |
| $B$ | 224 | 227 | $229^6$, $212^7$, $235^5$ | $175\ (5.2)^{36}$, $174\ (4.8)^{37}$, $174^{38}$ |
| $G$ | 72 | 75 | - | $74^{39}$ |
| $Y$ | 194 | 203 | - | $196^2$, $177^{40}$, $200^{39}$, $200^{41}$ |
| $\upsilon$ | 0.36 | 0.35 | - | $0.36^{39}$ |

**TABLE VI.** The shear and compressibility anisotropy factors of cementite obtained from first-principles calculations. The linear bulk moduli (in GPa) along $a$, $b$, and $c$ axes, defined as $B_i = i\,dP/di$ ($i=a, b, c$) are also shown.

| Method | $A_1$ | $A_2$ | $A_3$ | $A_G$(%) | $A_B$(%) | $B_a$ | $B_b$ | $B_c$ |
|---|---|---|---|---|---|---|---|---|
| Energy-strain | 0.17 | 1.40 | 1.27 | 32.2 | 0.24 | 807 | 649 | 588 |
| Stress-strain | 0.21 | 1.40 | 1.27 | 27.4 | 0.29 | 838 | 645 | 596 |

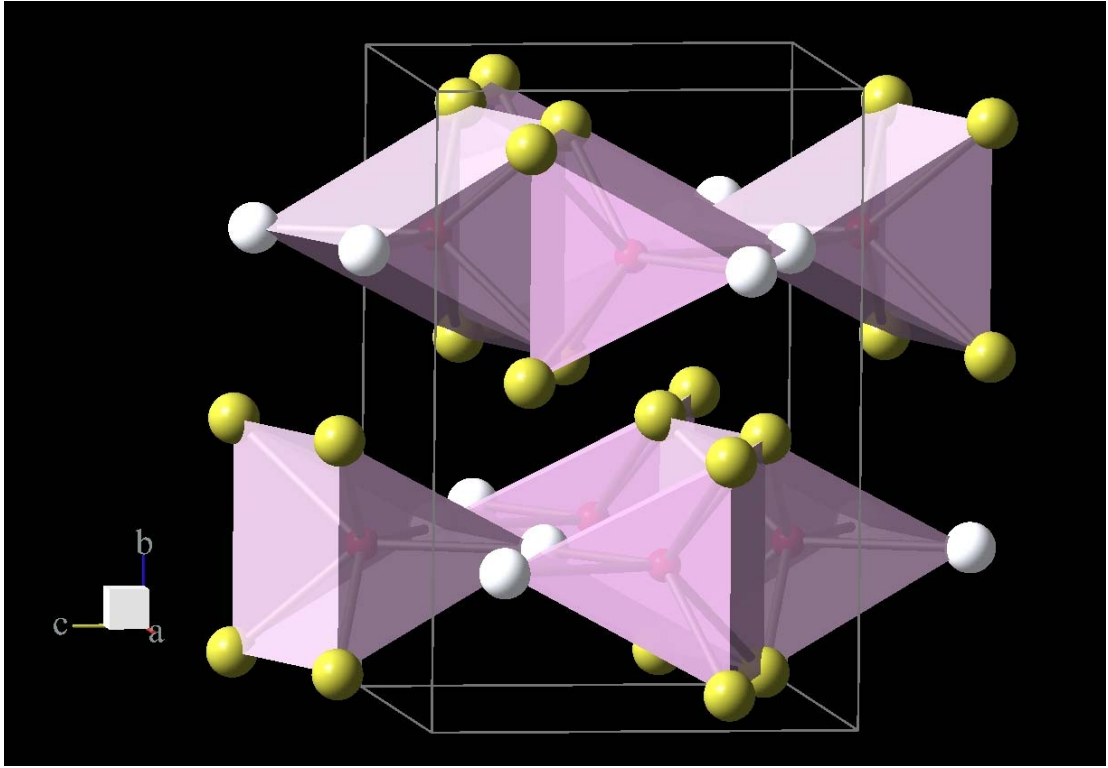

**FIG. 1**. Equilibrium crystal structure of cementite (Fe$_3$C). White, yellow, and red spheres represent Fe$_I$, Fe$_{II}$, and C atoms, respectively. All nearest-neighbor Fe-C bonds are also shown.

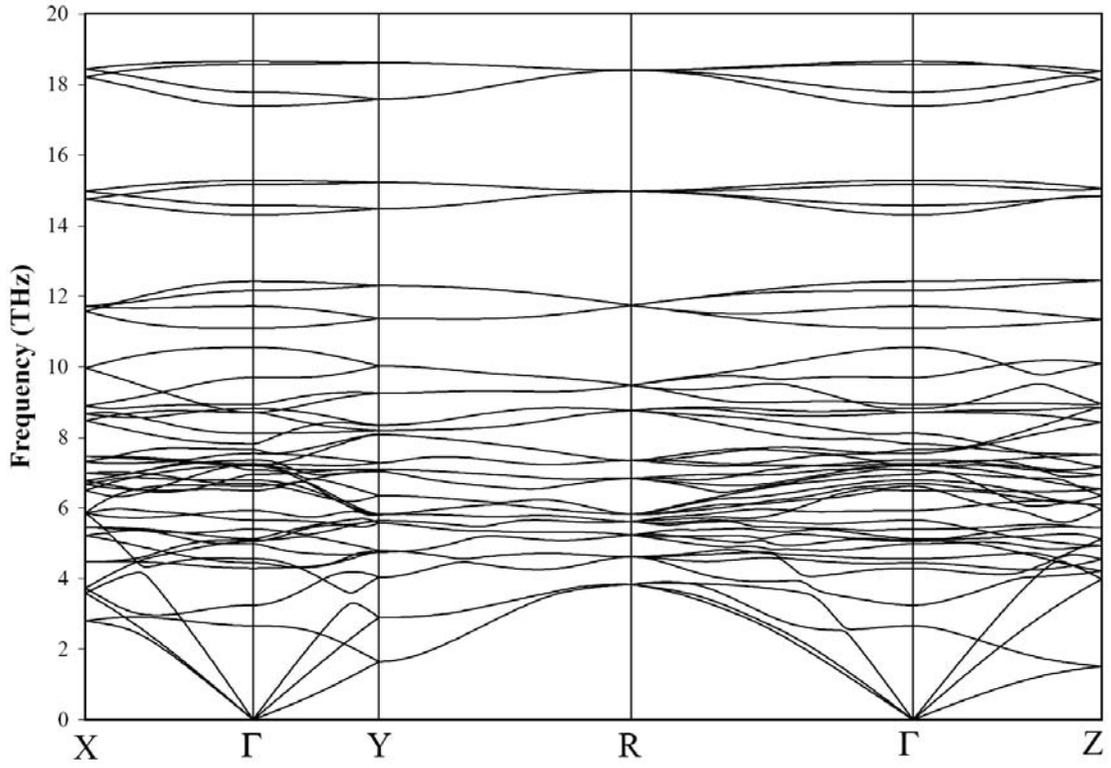

(a)

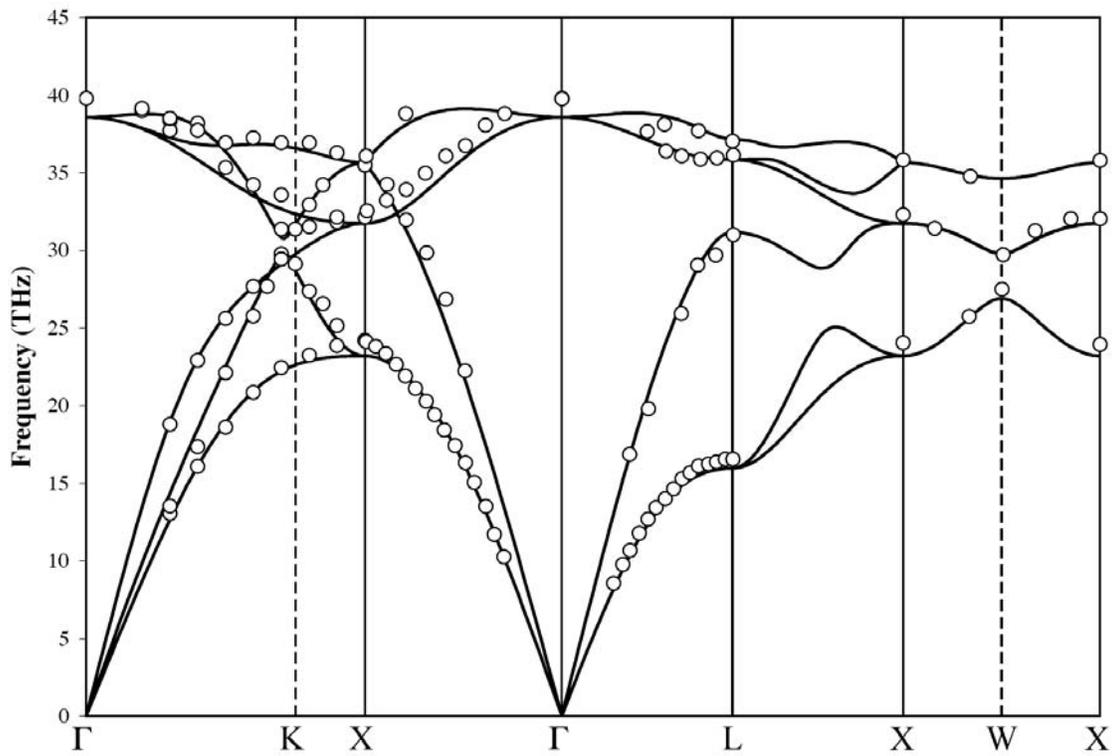

(b)

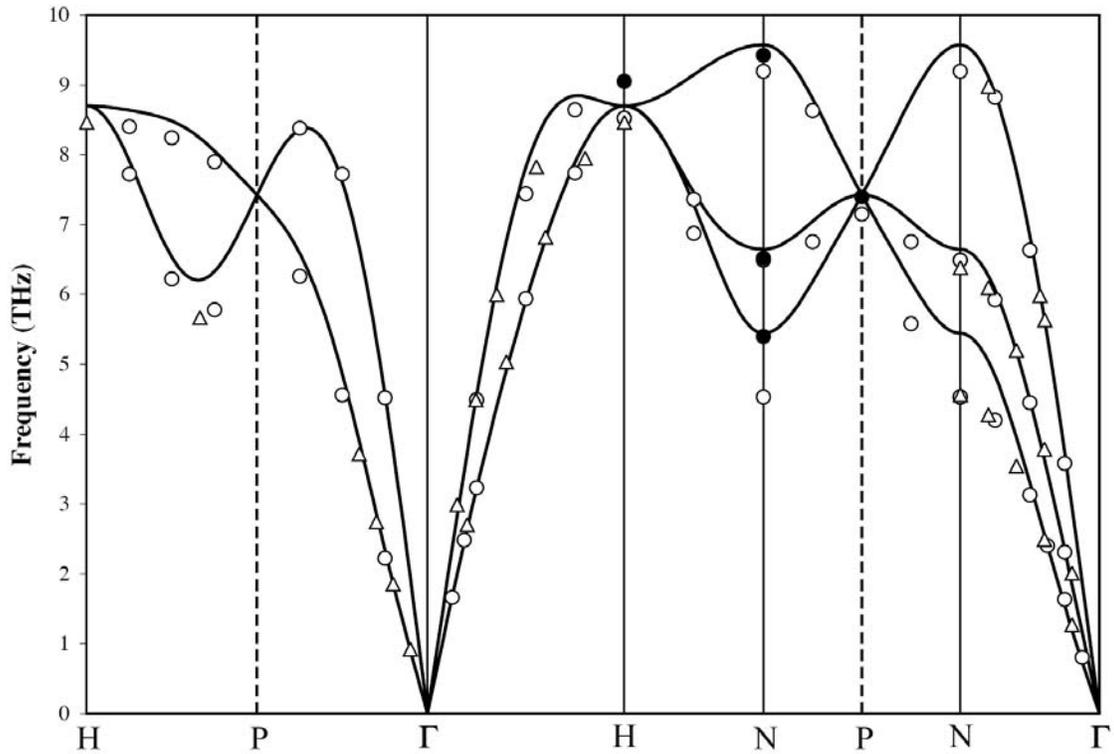

(c)

**FIG. 2**. Phonon dispersion relations for (a) cementite, (b) diamond, and (c) bcc Fe along various high-symmetry directions in the Brillouin zone calculated using 128-atom supercells. For diamond, experimental data (circles) are taken from Warren et al.[31] For bcc Fe, experimental data are taken from Brockhouse et al.[32] (circles) and Klotz and Braden[33] (triangles). Results from frozen-phonon calculations are shown as filled circles.

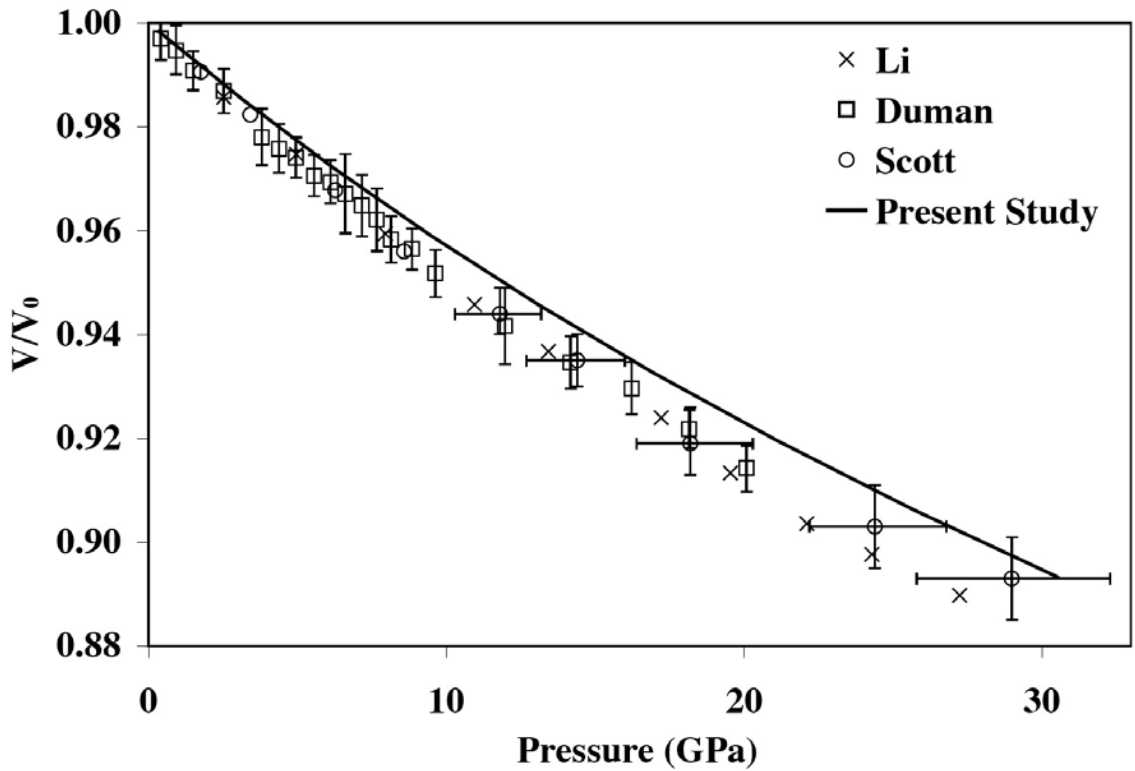

**FIG. 3**. First-principles calculated equation of state of cementite in comparison with experimental measurements. The errors bars for the experimental data from Li *et al.*[37] are not available.

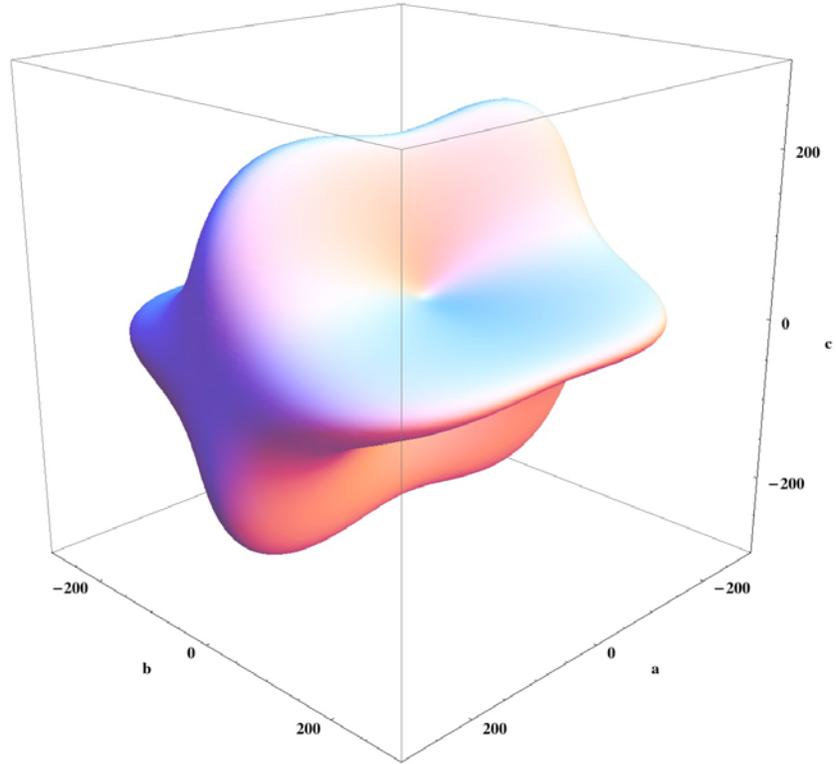

**FIG. 4**. Representational surface showing the directional dependence of the Young's modulus (in GPa) in cementite.

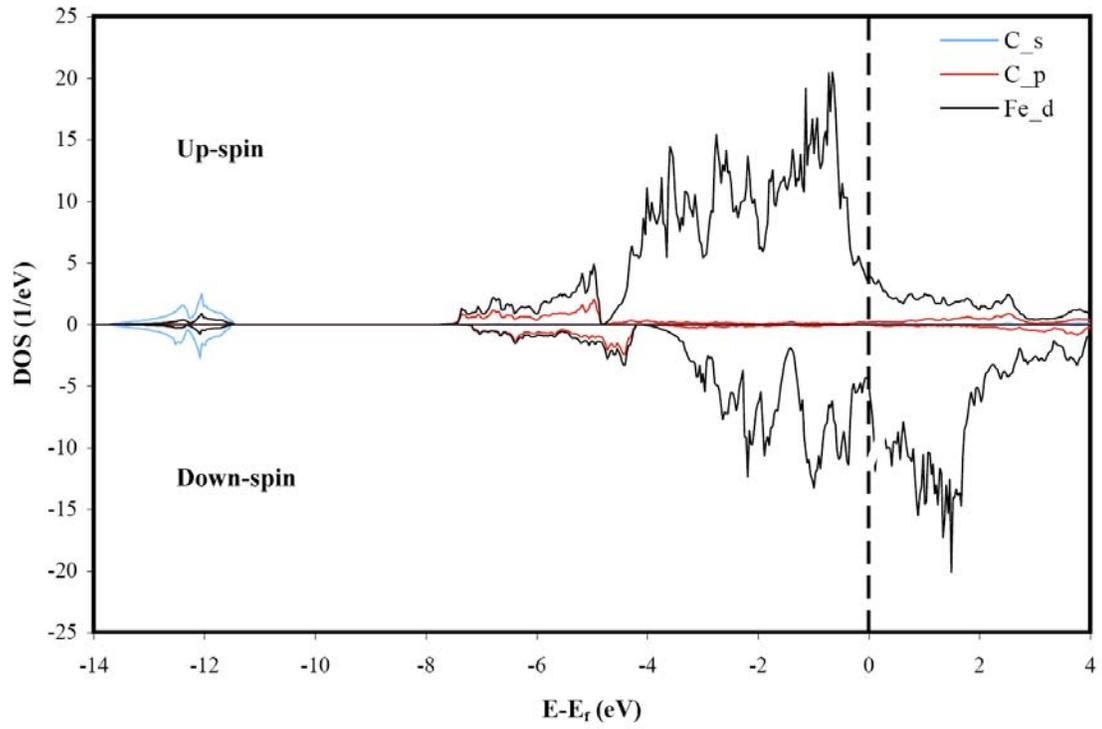

**FIG. 5**. Site and angular momentum projected electronic density of states of cementite. The vertical line denotes the Fermi level.

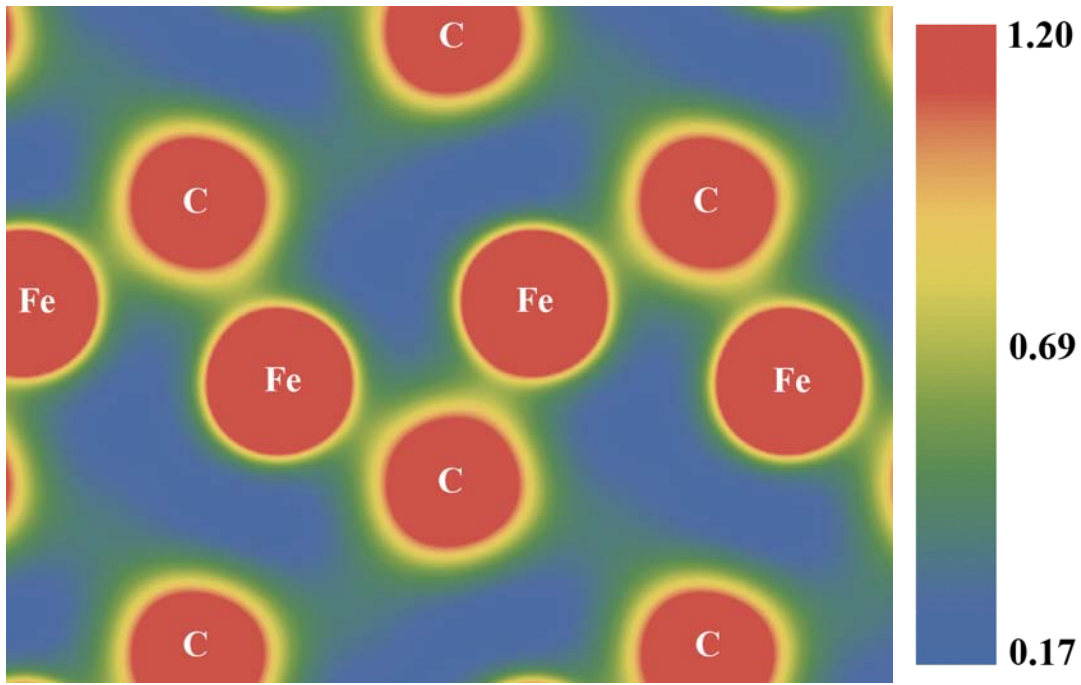

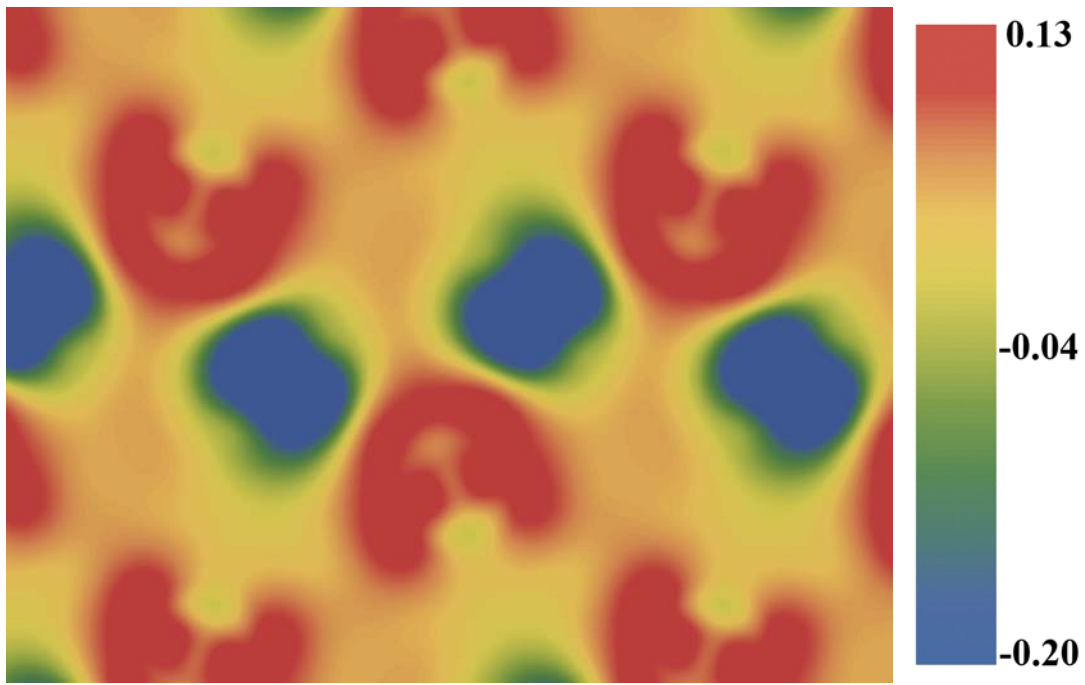

**FIG. 6**. Plot of (a) total charge density and (b) difference charge density, all in e/Å$^3$, on the (010) plane of Fe$_3$C containing the nearest-neighbor Fe-C bonds.